\documentclass[pra,usletter,superscriptaddress,twocolumn]{revtex4}
\usepackage{amsmath,latexsym,amssymb,amsfonts}

\begin{document}

\preprint{} \title{Detecting multipartite entanglement} 
\author{  Andrew C. Doherty} 
\affiliation{School of Physical Sciences,
  University of Queensland, St Lucia, Queensland 4072, Australia}
\affiliation{Institute for Quantum Information, California Institute of
  Technology, Pasadena,CA 91125, USA} 
\author{Pablo A. Parrilo}
\affiliation{Institut f\"ur Automatik, ETH Zentrum, CH-8092 Z\"urich,
  Switzerland} 
\author{Federico M. Spedalieri} 
\affiliation{Quantum Computing Technologies Group, Jet Propulsion Laboratory,
Pasadena, 91109, USA}
\affiliation{Institute  for Quantum Information, California Institute
  of Technology, Pasadena,CA 91125, USA}

\date{\today}

\begin{abstract}
  We discuss the problem of determining whether the state of several
  quantum mechanical subsystems is entangled. As in previous work on
  two subsystems we introduce a procedure for checking separability
  that is based on finding state extensions with appropriate
  properties and may be implemented as a semidefinite program. The
  main result of this work is to show that there is a series of tests
  of this kind such that if a multiparty state is entangled this will
  eventually be
  detected by one of the tests. The procedure also provides a means of
  constructing entanglement witnesses that could in principle be
  measured in order to demonstrate that the state is entangled.
\end{abstract}
 \maketitle

\section{Introduction}

Entanglement has long been recognised as one of the central features
of quantum mechanics and has been a primary focus of research in
quantum information science over recent years because of its central
role in phenomena such as teleportation, quantum cryptography and
violation of Bell inequalities \cite{nielsen2000}. A common theme of
theoretical research is the notion that entanglement is a resource
that often makes it possible to acomplish tasks that cannot be
performed in analogous classical scenarios. However, much of this
intuition is based on our theoretical understanding of pure states of
two separated systems. For mixed states and for states of many
separated systems much less is known. In this paper we address the
question of how to determine whether a given mixed state of
several subsystems is entangled.

Entangled states of separated quantum systems, atoms or photons
for example, are those that cannot be prepared by local operations and
classical communication. In order to prepare entangled states it is
necessary to have a non-trivial coherent interaction between the
different subsystems. As a result a state $\rho$ of $N$  subsystems
defined in $\bigotimes_{i=1}^{N} {\cal H}_{A_i}$,
is said to be \emph{fully separable} \cite{werner1989a}, that is not
entangled, if it can be written as
\begin{equation}
\label{sep}
\rho =\sum_i p_{i} \bigotimes_{j=1}^N |\psi^{(A_j)}_{i}\rangle \langle
\psi^{(A_j)} _{i}|,
\end{equation}
where the $|\psi^{(A_j)}_{i}\rangle$ are state-vectors on the spaces
${\mathcal{H}}_{A_j}$ and $p_i >0, \sum_i p_i =1$. If such a
decomposition does not exist, the state cannot be prepared by local
operations and classical communication between the parties and is
termed entangled. The so-called {\em
  separability problem} arises from the fact that even for the case of
two parties, and even given complete information about the matrix
elements of the density operator of the system, it is difficult to
determine whether such a decomposition as a mixture of product pure
states exists. Much of the difficulty arises because density matrices
can generally be decomposed into many different ensembles of pure states. 


The separability problem for bipartite systems has received much
attention and we refer the reader to one of the several reviews
\cite{terhal2002a,lewenstein2000b,bruss2002a}. However, as a result of
recent work by Gurvits on the computational complexity of the problem
\cite{gurvits2002a} it is extremely unlikely that any completely
satisfactory solution can exist. Since Gurvits showed that the
separability problem
for a given bipartite mixed state is
in the complexity class NP-HARD, it is extremely unlikely that any
algorithm that checks whether a quantum state is entangled can be
performed with an amount of computation that is polynomial in the
dimension of the Hilbert spaces involved.  

The worst case complexity of the problem is not the end of the story.
There are simple, efficiently computable, tests that can establish the
entanglement of a large subset of states. The most well-known of these
is the positive partial transpose or Peres-Horodecki criterion
\cite{peres1996a,horodecki1996a,woronowicz1976a}. This
simply requires making an appropriate rearrangement of the matrix
elements of $\rho$, corresponding to transposing one of the parties,
and checking that the resulting matrix is positive. In
\cite{doherty2002a} we proposed a
hierarchy of separability criteria that can be thought of as a
generalisation of this
condition but which can only be checked by solving a semidefinite
program. We subsequently showed, based on earlier work
\cite{fannes1988a,raggio1989a}, that this series of tests
was complete in the sense that any entangled state of two subsystems
would eventually be detected by one of the tests in our hierarchy
\cite{doherty2004a}.
Another attractive feature of these conditions is that if a given test
successfully identifies that the state of interest is entangled it also
constructs an observable, known as an entanglement witness, that could
in principle be measured in order to demonstrate this entanglement
experimentally.

Semidefinite programs are members of a class of convex optimizations
that may be solved with arbitrary accuracy in polynomial time
\cite{VaBbook,VaB:96}. By identifying the separability criteria in
\cite{doherty2002a,doherty2004a} as semidefinite programs it was
possible to assess the computational difficulty of the criteria and to
construct entanglement witnesses when the criteria successfully
determine that a given state is entangled. 
Techniques from convex optimization are being applied increasingly
frequently in quantum information, notable examples include
\cite{jonathan1999a,rains2001a,audenaert2001b,verstraete2003a,kitaev2004a}.

\subsection{Overview of results and relation to other work}

In this paper we extend our results to the case of an arbitrary number
of parties. One might think that an approach to the problem of
determining whether a given
multiparty state is entangled would be to consider the different ways
that the subsystems can be collected into two groups and determine
whether the resulting bipartite states are entangled. In fact, all pure
entangled states will result in a bipartite entangled state for some
grouping into two parties. As a result, the reduced density matrix for
some subset of the systems will have non-zero entropy thus showing
that the state is entangled. Indeed 
checking any reduced density matrix will suffice for generic
states. It is clearly
possible to determine whether or not a pure state is
entangled in this 
way regardless of what class of multipartite pure entangled states,
such as the GHZ and W states of three qubits \cite{duer2000b}, are
considered. For mixed states, however, no such solution is
possible. There are entangled states that
are separable whenever the parties are arranged into two groups, as
was first shown by an example constructed from unextendible product
bases \cite{bennett1999a}. The
multipartite separability problem cannot be reduced to a series of
bipartite separability problems. In general it is possible to classify
states based on their separability when the $N$ particles are grouped
into any number $k\leq N$ of groups. This classification was developed
in detail by D\"{u}r and collaborators \cite{duer1999a,duer2000a}. 

Despite the extra difficulty of the multiparty case much of the
structure of the bipartite separability problem is unchanged. There is
a nice discussion of work on the multiparty separability problem in
the review by Terhal \cite{terhal2002a}. A particularly important
observation is that the set of fully separable states forms a compact
convex set in the state space. In the bipartite case the separating
hyperplane theorem of convex analysis guarantees that a state is
entangled if and only if there is an observable known as an
entanglement witness that detects this entanglement
\cite{woronowicz1976a,horodecki1996a}. Entanglement
witnesses are observables that have a positive expectation value for
every separable state and a negative expectation value for some
entangled state. Just as in the bipartite case the separating
hyperplane theorem guarantees that if a multiparty state $\rho$ is
entangled then there is an observable $W$ with a negative expectation
value $\mathrm{Tr}[W\rho]<0$ but a positive expectation value for all
fully separable states \cite{horodecki2000b}. This convexity structure
and the resulting entanglement witnesses exactly mirror the bipartite
separability problem.

Checking whether $\rho$ is separable is equivalent to checking
whether an entanglement witness exists. In the bipartite case this
reduces to a problem that may be stated in terms of polynomial
inequalities since entanglement witnesses map onto positive
semidefinite bihermitian forms. The multipartite
separability problem may
still be phrased as quantified polynomial inequalities:
\begin{equation}
\label{entopt}
\forall W \ [ 
\forall P_{\mathrm{prod}} \ \mathrm{Tr} [P_{prod} W] \geq 0 
\Longrightarrow
 {\mathrm{Tr}}[\rho W] \geq 0 
],
\end{equation}
where $P_{prod}=\bigotimes_j P_j$ is a pure product state and
$P_j=|\psi^{(A_j)}\rangle \langle \psi^{(A_j)}|$ a rank one projector
on ${\cal H}_{A_i}$.  By writing the condition $\mathrm{Tr} [P_{prod}
W] \geq 0$ in terms of the
components of the various $|\psi^{(A_j)}\rangle$ it is clear that the
polynomials that arise in the multipartite case 
are no longer bihermitian but multihermitian; that is hermitian in the
sets of variables corresponding to each of the subsystems. If this
proposition is satisfied then $\rho$ is 
fully separable. 

Problems that may be written in terms quantified
polynomials inequalities of a finite number of variables (the components of $W,
|\psi^{(A_j)}\rangle $) are known as semi-algebraic problems.
Semi-algebraic problems are known to be \emph{decidable} by the
Tarski-Seidenberg decision procedure \cite{BCR98} which provides an
explicit algorithm to solve the separability problem in all cases and
therefore to decide whether $\rho$ is entangled. Exactly the same is
true of the bipartite problem but as we noted in \cite{doherty2004a},
exact techniques in
algebraic geometry that could be used to solve the separability
problem scale very poorly with the number of variables and tend not to
perform well in practice except for very small problem instances. Such
general methods of algebraic geometry have, however, been applied to
the separability problem \cite{ioannou2004a} and related problems
\cite{jamiolkowski1974a}.

As we noted above there are efficient procedures that, like the PPT
test, demonstrate that a state is entangled in many cases. In general,
algorithms that are able to solve in polynomial time 
many but not all problem instances of a computationally hard problem
are not excluded 
by complexity theory (even presuming that $P\neq NP$). In fact in
\cite{parrilo2003a} one
of us showed that for all semi-algebraic problems it is possible to
construct a series of semidefinite programs that are able to solve
large classes of problem instances.  A direct application of those
techniques would lead to a complete hierarchy of efficiently
computable separability criteria such that every entangled state would
be detected at some level in the hierarchy. However, the most obvious
version of this would result from writing all the variables and
parameters (the state, the coefficients of
$W$ and so on) in terms of their real and imaginary parts and treating
the resulting problem involving real polynomials as a question in real
algebraic geometry, to which the methods of \cite{parrilo2003a}
apply directly. The resulting sequences of criteria would be difficult
to interpret in terms of the original quantum mechanical problem
structure. In this paper we show how to construct a complete series of
multiparty separability criteria that, while falling in the general
scheme of \cite{parrilo2003a}, may be phrased directly in terms of
quantum mechanical states and observables.

A recent series of papers has considered a slightly different setting
for both the bipartite and multipartite separability
problems \cite{brandao2004a,brandao2004b,brandao2004c}. Brand\~{a}o
and Vianna point out that the separability
problem is an example of a class of convex optimizations known as
robust semidefinite programs. Although robust semidefinite programs,
just as semi-algebraic problems,
are computationally difficult there are also well studied relaxations that
are able to address certain problem instances. Brand\~{a}o and Vianna
show that both deterministic algorithms that are able to solve some
problem instances \cite{brandao2004a} and probabilistic algorithms
that give correct
answers with some probability \cite{brandao2004b} can provide
tractable approaches to the 
separability problem, at least in low dimensions. 

The tests we consider are an obvious generalisation of
\cite{doherty2004a} to the multiparty case. As such they revolve
around the question of whether certain symmetric state extensions
exist for $\rho$. The general problem of when a global state is
consistent with a given set of reduced density matrices for various
overlapping subsystems of a multipartite quantum system has a long
history. The importance of this general state extension problem was
emphasised by Werner in \cite{werner1989b,werner1990a}. A simple
example is to specify that two systems $A$ and $B$ are in some
entangled pure state and that $B$ and a third system $C$ have a
reduced density matrix that is also this same pure entangled state.
That this specification of reduced states is inconsistent with any
quantum state for the whole system $A,B,C$ is known in quantum
information as the monogamy of pure state entanglement
\cite{terhal2003b}; given that two quantum systems are in a pure
entangled state it is not possible for either one to be entangled with
a third system. Mixed
entangled states also tend to be monogamous; Werner used the violation
of Bell inequalities for certain mixed bipartite entangled states $\rho$ to
rule out the existence of a state of $A,B,C$ where the reduced states
of both $A,B$ and $C,B$ are $\rho$ \cite{werner1990a}. This logic can
be reversed; the existence of such a global state on $A,B,C$ implies
that there is a local hidden variable description for certain Bell
experiments on $\rho$ \cite{terhal2003a} and this construction can
readily be extended to multiparty cases 
\cite{terhal2003a,wolf2003a}. The connection between this consistency
problem for reduced states and the bipartite separability problem
which is central to \cite{doherty2002a,doherty2004a} is in fact
made in a brief comment in \cite{werner1989a}. Using the techniques of
\cite{doherty2002a,doherty2004a} all of the
state
extension problems resulting from specifying sets of reduced density
matrices and asking if this
specification is consistent with a global mixed state can be phrased
as semidefinite programs, a fact has
important implications for practical calculations. 

The question of when a specification of reduced states for a quantum
system is consistent with a global state for the system was raised
again in \cite{linden2002b}. Subsequent work has focussed on when a
set of 
one-party reduced density matrices is consistent with a \emph{pure} state of
the joint system for some number of qubits or
qutrits~\cite{higuchi2003a,bravyi2003a,higuchi2003b,han2004a}. The
situation when two-party reduced density matrices are specified for
mixed states of
three quantum systems is considered in \cite{han2004a}. In each of these cases
it is possible to derive necessary conditions for compatibility based
on the eigenvalues of the reduced density matrices. Very recent work
by Jones and Linden \cite{jones2004a} shows that the general question
of when a set of reduced states is consistent with a pure quantum
state for the whole system is
expressible as a specific problem in real algebraic geometry. This
seems to be a very significant difference to the 
version of the problem in which the joint state is allowed to be mixed
since most interesting classes of problems in real
algebraic geometry prove to be computationally hard while semidefinite
programs may be solved in polynomial time. See
\cite{parrilo2003a} for a discussion of this point and algorithms that
solve problems in real algebraic geometry using
semidefinite programming. In other
important recent progress on state extension problems, Linden and
Wootters \cite{linden2002a} have shown that the
reduced density matrices of a certain
fraction of the parties of a generic multi-party pure state completely
determine the state; the bounds on this fraction have been
significantly improved in \cite{jones2004a}.

Another very important instance of this state extension problem,
termed by Coleman the N-representability problem \cite{coleman1963a},
has been much
studied in physical chemistry over a long period (for recent
discussions and references see
\cite{coleman2000a,mazziotti2001a,mazziotti2002a}). The
N-representability problem poses
the question of which two-body reduced density matrices are consistent
with a valid global state of $N$ fermions. The antisymmetrization of
the fermion wavefunction requires that all two-particle reduced
density matrices be the same and the global state be antisymmetric to
swapping particles. The reason for interest in
this problem is that the ground state energy of an interacting fermion
system
can be written in 
terms of the two-body reduced density matrix if only two-body
interactions occur in the Hamiltonian. A lot of information about the
ground states of molecular systems could be found if tractable
conditions for N-representability existed. A similar connection
between state extension problems and the
ground states of spin systems with local interactions was also noted
by Werner \cite{werner1990a}. In the tradition
of work on this problem in physical chemistry necessary conditions for
N-representability are often found in terms of conditions on the
particle and hole correlations  and it has recently been
realised that these in turn may be able to be expressed as
semidefinite programs \cite{nakata2001a,mazziotti2002a,zhao2004a}.

The key idea of this paper is to propose a sequence of state extensions
that must exist if a given multiparty quantum state $\rho$ is
separable. Like all state extension problems these may be expressed as
semidefinite programs. The key result is the determination that this
sequence of tests is complete in the sense that it can in principle
detect all entangled states. This is achieved by an inductive argument
in the number of parties. Like \cite{doherty2004a} this argument
depends on the strengthened version of the quantum de Finetti theorem
proven in \cite{fannes1988a,raggio1989a}.

The rest of the paper is structured as follows. In
Sec.~\ref{extensions} we introduce the separability criteria we will
consider. As discussed in Sec.~\ref{entwits} these can be checked by
solving a semidefinite program and we show how to use the theory of
semidefinite programming to construct entanglement witnesses for
$\rho$ whenever one of the criteria shows $\rho$ to be entangled. The
central result that a given series of separability criteria is
complete in the sense that any entangled state will be detected by
some test in the series is proven in Sec.~\ref{completeness}. In
Sec.~\ref{example} we explicitly consider the example of Bennett {\it
  et al.} \cite{bennett1999a} of a completely bound entangled state
where no PPT test or bipartite separability test would suffice to
demonstrate that the state is entangled. Finally, we conclude in
Sec.~\ref{conclude}. 

\section{Multipartite separability criteria}
\label{extensions}
Let $\rho$ be a N-partite state defined in $\bigotimes_{i=1}^{N} {\cal
  H}_{A_i}$, where the different parties $A_i$ are represented by
Hilbert spaces ${\cal H}_{A_i}$ of dimension $d_{A_i}$ respectively.
Let $\vec{n} = (n_1,\ldots,n_N)$ be a vector of positive integers
greater than or equal to one. We will say that a state
  $\rho_{\vec{n}}$ defined in
$\bigotimes_{i=1}^{N} {\cal H}_{A_i}^{\otimes n_i}$, which can be
viewed as the original space supplemented by $(n_i -1)$ copies of
party $A_i$, is a \emph{locally symmetric extension (LSE) of} $\rho$,
if it satisfies the following two properties:
\begin{enumerate}
\item $\rho_{\vec{n}} = V_{i,\tau (i)} \, \rho_{\vec{n}} V_{i,\tau
    (i)} \ \ \forall i, 1\leq i \leq N$, and $\forall \tau (i) \in
  S_{n_i}$, with
\begin{equation}
 V_{i,\tau (i)} = \left(\bigotimes_{j=1}^{i-1} \openone_{A_j}^{\otimes n_j}\right)
\otimes \Pi_{\tau (i)} \otimes  \left(\bigotimes_{j=i+1}^{N} \openone_{A_j}^{\otimes n_j}\right),
\end{equation}
where $S_{n_i}$ is the group of permutations of $n_i$ objects and $
\Pi_{\tau (i)}$ is the operator that applies the permutation
$\tau (i) \in S_{n_i}$ to the $n_i$ copies of party $A_i$.
\item $\rho = \mathrm{Tr}_{\{A_1^{\otimes n_1 -1} \cdots A_N^{\otimes
      n_N -1}\} } [ \rho_{\vec{n}}]$.
\end{enumerate}
The first property means that $\rho_{\vec{n}}$ remains invariant
whenever we permute the copies of a certain party. Due to this symmetry,
we do not need to specify which copies of $A_i$ are we tracing over in
the second property. Furthermore, we can define a PPT locally
symmetric extension (PPTLSE), by requiring $\rho_{\vec{n}}$ to remain
positive semidefinite under any possible partial transposition.

We will now show that we can use this definition to generate a family
of separability criteria. It is very easy to see that any fully
separable state has LSE for any vector $\vec{n}$. This can be seen
from (\ref{sep}), since the state
\begin{equation}
\label{lse}
\rho_{\vec{n}} =\sum p_{i} \bigotimes_{j=1}^N \left(|\psi^{(A_j)}_{i}\rangle 
\langle \psi^{(A_j)} _{i}|\right)^{\otimes n_i},
\end{equation} 
clearly has the required properties. Moreover, the state in
(\ref{lse}) is obviously PPT, since it is fully separable. We have
then the property that any fully separable state has PPTLSE to any
number of copies of its parties.  This observation can be used to
generate a family of separability criteria. Any state that fails to
have a PPTLSE for some number of copies \emph{must be entangled}.

For any vector $\vec{n}$ that represents the number of copies of the
different parties, we can construct a separability criterion by just
asking the question of whether the state $\rho$ has a PPTLSE to that
particular number of copies.  Thus, we can construct a countably
infinite family of separability criteria.  This is similar to the
situation in the bipartite case discussed in~\cite{doherty2004a}.
However, in the multipartite case, these criteria cannot be all
ordered in a hierarchical structure, although they have a natural
\emph{partial order}. For example, if a state has a PPTLSE to
$\vec{n}$ copies, then it clearly has PPTLSE to $\vec{k}$ copies, for
all $\vec{k}$ that satisfy $k_i \leq n_i, \forall i $, since we can
construct such an extension by tracing $(n_i - k_i)$ copies of party
$A_i$, $1\leq i \leq N$.  This property of the extensions is mapped
into the partial order of N-tuples given by
\begin{equation}
\label{parord}
\vec{k} \preceq \vec{n} \iff k_i \leq n_i , \forall i, 1\leq i\leq N.
\end{equation}
Conversely, if a state does not have a PPTLSE to $\vec{k}$ copies,
which means it is entangled, then it cannot have PPTLSE to $\vec{n}$
copies, for any $\vec{n}$ satisfying $\vec{k} \prec \vec{n}$. However,
there does not seem to be any relationship between the existence of
PPTLSE to number of copies whose vectors are not related by the
partial order (\ref{parord}).

In the following section we will discuss the semidefinite programs
that determine whether a state has a PPTLSE. By using the duality
theory of semidefinite programs we will show how to construct
entanglement witnesses in cases where a PPTLSE fails to exist. 

\section{Separability criteria as semidefinite programs and
  entanglement witnesses} 
\label{entwits}
The techniques of \cite{doherty2002a,doherty2004a} allow us to determine
whether a given PPTSE exists by solving a semidefinite programming
feasibility problem. Such problems amount to deciding whether there
exists a positive matrix subject to given affine constraints. We will
not dwell on the details here which are essentially identical to
\cite{doherty2002a,doherty2004a}.

We begin by noting that the state extension (\ref{lse}) satisfies a
stronger property than invariance under swapping the copies of the
different Hilbert spaces. Let us denote the symmetric subspace of $k$
copies of ${\cal H}_{A_i}$ by $\mathrm{Sym}^k(A_i)$. Let $\pi_k^{(i)}$
be the projectors onto these subspaces. Then the PPTLSE of Eq.\ 
(\ref{lse}) has support on the tensor product of these symmetric
subspaces $ \mathrm{Sym}^{n_1}( A_1)\otimes \cdots \otimes
\mathrm{Sym}^{n_N}(A_N)$. For all $i$ the PPTLSE of Eq. (\ref{lse})
satisfies $\pi_{n_i}^{(i)} \rho_{\vec{n}}\pi_{n_i}^{(i)}
=\rho_{\vec{n}}$. More economically we may define a projector
$\pi_{\vec{n}} = \prod_i \pi_{n_i}^{(i)}$ onto the subspace $\bigotimes_i
\mathrm{Sym}^{n_i}(A_i)$.

Since the extension must remain positive under all possible partial
transpositions, we need to impose a whole set of positivity
constraints on $\rho_{\vec{n}}$.  We will write then
\begin{equation}
\rho_{\vec{n}}^{T_{\cal S}}  \geq 0,
\end{equation}
where we use ${\cal S}$ to represent any subset of the tensor factors
in $\bigotimes_{i=1}^{N} {\cal H}_{A_i}^{\otimes n_i}$ that yields an
independent partial transpose, including the empty set, which we will
associate with not applying any partial transposition.

To summarize the conditions on $\rho_{\vec{n}}$, for a given $\vec{n}$
we must
\begin{eqnarray}
\label{sdp}
{\mathrm{find}} &\ \  \rho_{\vec{n}}  \nonumber \\
{\mathrm{ subject \ to}} &\ \  \rho_{\vec{n}}^{T_{\cal S}} \geq 0
\quad \forall {\cal S} \nonumber \\
&\ \ \pi_{\vec{n}} \rho_{\vec{n}}\pi_{\vec{n}} 
=\rho_{\vec{n}}  \nonumber \\
&\ \ \mathrm{Tr}_{\{A_1^{\otimes n_1 -1} \cdots A_N^{\otimes n_N -1}\} } [ \rho_{\vec{n}}]=\rho. 
\end{eqnarray}
Both of the equalities above can be written in terms of a finite
number of trace constraints by writing them in terms of an explicit
basis for Hermitian matrices as in \cite{doherty2004a}. So the partial
trace conditions on $\rho_{\vec{n}}$ define an
affine subset of matrices on $\bigotimes_{i=1}^{N} {\cal
  H}_{A_i}^{\otimes n_i}$ and if a positive symmetric state extension
exists this subset will intersect with the cone
of positive semidefinite matrices. Determining whether the intersection is
empty is a semidefinite programming feasibility problem.

We may now apply the duality theory of semidefinite programs to find
the dual optimization \cite{VaBbook}. This optimization proves to be a
search for an
entanglement witness $Z$. The dual optimization is
\begin{eqnarray}
\label{dual}
{\mathrm{minimize}} &\ \  \mathrm{Tr}[Z \rho]  \nonumber \\
{\mathrm{ subject \ to}} 
&\ \ Z_{\mathcal S} \geq 0 \quad \forall {\mathcal S}
 \nonumber \\ 
&\ \  \pi_{\vec{n}} \left( Z\otimes I  \right)\pi_{\vec{n}} = \pi_{\vec{n}}
\left( \sum_{\mathcal S} Z_{\mathcal S}^{T_{\mathcal S}}\right) \pi_{\vec{n}} 
\end{eqnarray}
Note that $Z$ is an observable on the physical Hilbert space
$\bigotimes_i {\cal H}_{A_i} $ and the identity $I$ acts on the
duplicate copies of different parties $\bigotimes_{i=1}^{N} {\cal
  H}_{A_i}^{\otimes (n_i-1)}$. Thus the different $Z_{\mathcal S}$ are
observables on the same space as the state extensions
$\rho_{\vec{n}}$, $\bigotimes_{i=1}^{N} {\cal H}_{A_i}^{\otimes n_i}$.
We will show that this dual optimization answers the question of the
existence of a PPTLSE $\rho_{\vec{n}}$ equally well and has the added
benefit that when such an extension does not exist the optimum $Z^*$
is an entanglement witness.

Suppose that some $Z^*$ satisfying these constraints exists and has
Tr$[Z^*\rho]<0$ and yet there is also a PPTLSE $\rho_{\vec{n}}$. Then
\begin{eqnarray*}
\mathrm{Tr}[Z^*\rho]&=&\mathrm{Tr}\left[\left( Z^*\otimes I \right)
  \rho_{\vec{n}}\right] \\ & =&
  \mathrm{Tr} \left[ \pi_{\vec{n}}
\left( \sum_{\mathcal S} Z_{\mathcal S}^{T_{\mathcal S}}\right)
  \pi_{\vec{n}} \rho_{\vec{n}}\right] \\
&=& \sum_{\mathcal S}  \mathrm{Tr} 
\left[Z_{\mathcal S}
  \rho_{\vec{n}} ^{T_{\mathcal S}}\right] \\
&\geq& 0,
\end{eqnarray*}
which is a contradiction. The first line follows from the fact that
$\rho_{\vec{n}}$ is an extension for $\rho$, the second line from the
symmetry of $\rho$ and the constraints on $Z$. The third again uses
the symmetry of $\rho$ and the property of partial transposes that
Tr$[X^{T_{\mathcal S}}Y]=\mathrm{Tr}[XY^{T_{\mathcal S}}]$. Finally
positivity results from the requirement that both $\rho_{\vec{n}}$ and
the different $Z_{\mathcal S}$ are positive semidefinite. If such an
observable $Z^*$ exists then $\rho$ cannot have a PPTLSE
$\rho_{\vec{n}}$ and thus $\rho$ must be entangled. Equally all
separable states $\sigma$ do have have a PPTLSE $\sigma_{\vec{n}}$
given by Eq.\ (\ref{lse}) and as a result Tr$[Z^*\sigma] \geq 0$.
Therefore $Z^*$ is an entanglement witness. As in the bipartite case
discussed in \cite{doherty2004a} these entanglement witnesses have
interesting algebraic properties that relate them to the general
methods of \cite{parrilo2003a}. Since the details are essentialy
identical to the bipartite case we refer the interested reader to
these two references.

This leaves the possibility that the optimum of the dual semidefinite
program (\ref{dual}) is positive and yet no PPTLSE $\rho_{\vec{n}}$
exists. As in the bipartite case this possibility must be excluded by
appealing to strong duality \cite{doherty2004a}. Broadly speaking when
no PPTLSE exists the existence of an entanglement witness of the form
$Z^*$ is guaranteed by the separating hyperplane theorem of convex
analysis applied to an appropriate convex set associtated with the
feasibility problem (\ref{sdp}). However, in order to apply this
theorem we must check that this set is in fact closed. In our case
this may be determined by checking that $Z=I>0$ satisfies the
constraints of the dual semidefinite program (\ref{dual}). For full
details of this argument see Appendix B of \cite{doherty2004a}. We may
conclude that when no PPTLSE exists we may use the dual semidefinite
program to construct an entanglement witness and equally that the
optimum of the dual program can only be positive if a PPTLSE exists.

These two equivalent semidefinite programs can be implemented
numerically using exactly the techniques described in
\cite{doherty2004a} and we will not dwell on these details here. Once
again it is important to implement the optimizations in a way that
preserves the symmetries, making use of the fact that $\rho_{\vec{n}}$
can be restricted to lie on the symmetric subspace $\bigotimes_i
\mathrm{Sym}^{n_i}(A_i)$. For a fixed number of parties and a fixed
$\vec{n}$ the computation required to solve the two semidefinite
programs will scale polynomially with the Hilbert space dimensions
involved. Also for a fixed number of parties and fixed Hilbert space
dimensions the computation required to perform the tests will scale as
some polynomial of the components of $\vec{n}$. In this case the
number of inequivalent partial transposes will be limited very greatly
by the symmetry between the different copies of the subspaces $A_i$.
Unfortunately as the number of parties increases the number of
inequivalent partial transpose tests will increase very rapidly.
However the tests will be of use even if only a restricted subset of
the possible partial transposes (a restricted subset of the possible
${\mathcal S}$ in the above formulae) are actually used. The number of
inequivalent partial transposes is related to the number of possible
partitions of $N$ quantum systems and is discussed in
\cite{duer2000a}.

\section{Completeness of the family of tests}
\label{completeness}
Each test described in the previous section gives a necessary
condition for separability of a multipartite state. We have discussed
how these tests can be stated as semidefinite programs, which implies
that there are efficient algorithms to solve them. In this section we
will show that this family of criteria is also complete, in the sense
that any mutipartite entangled state will be detected by some test. We
will actually prove a stronger result; a weaker
family of tests is already complete. The proof is based on the
completeness of the bipartite hierarchy of
tests~\cite{raggio1989a,fannes1988a,doherty2004a}, and the properties
of the Quantum de Finetti representation~\cite{caves2002a}.

\newtheorem{thm1}{Theorem}
\begin{thm1}[Multipartite Completeness]
  Let $\rho$ be a multipartite mixed state in $\bigotimes_{i=1}^N
  {\cal H}_{A_i}$, such that $\rho$ has locally symmetric extensions
  (LSE) $\rho_{\vec{n}_k}$ for its first $(N-1)$ parties, associated
  with the vectors $\vec{n_k} = (k,k,\ldots,k,1), \, \forall k\geq 1$.
  Then $\rho$ is fully separable.
  
  Moreover, there are unique conditional probability densities
  $P_l(\omega_{A_l} | \omega_{A_{l-1}},\ldots,\omega_{A_1}), \, 1\leq
  l \leq N-1$, and a unique function $\lambda :
  D_{A_1}\times\cdots\times D_{A_{N-1}}\rightarrow D_{A_N}$, where $
  D_{A_i}$ is the space of states in ${\cal H}_{A_i}$, such that
\begin{eqnarray}
\rho & = & \int_{D_1^{N-1}}  \left( \bigotimes_{i=1}^{N-1}\omega_{A_i}
\right) \otimes 
\lambda_{A_{N}} (\omega_{A_1},\ldots,\omega_{A_{N-1}}) \times \nonumber \\
& & \times \ \Pi_{i=1}^{N-1} P_{i}(\omega_{A_{i}} | \omega_{A_{i-1}},\ldots,\omega_{A_1})
\ d\omega_{A_i},
\end{eqnarray} 
(with $\int_{D_1^{N-1}} $ meaning $ \int_{D_{A_1}} \cdots
\int_{D_{A_{N-1}}}$).
\end{thm1}
\textbf{Proof:} The proof is by induction in the number of parties.
The proof of the case $N=2$ is Theorem 1 in~\cite{doherty2004a}.

Let us assume the result holds for $N-1$. Let $\rho$ in
$\bigotimes_{i=1}^{N} {\cal H}_{A_i}$ have the LSE mentioned in the
statement of the theorem. Consider the split $A_1 -
(A_2,\ldots,A_{N})$ of the $N$ parties and regard $\rho$ as a
bipartite state.  Consider the LSE of $\rho$ associated with the
vector $\vec{n_k}$. Then, by tracing out $(k-1)$ copies of ${\cal
  H}_{A_i}$, $2\leq i \leq N$, we obtain a state in ${\cal
  H}_{A_1}^{k} \otimes (\bigotimes_{i=2}^N {\cal H}_{A_i})$ that is
invariant under permutations of the copies of ${\cal H}_{A_1}$ and
yields $\rho$ when we trace $(k-1)$ copies of ${\cal H}_{A_1}$. Hence,
$\rho$ has SE to any number of copies of $A_1$, and applying the
result of the bipartite case, we can write
\begin{equation}
\label{rho1}
\rho = \int_{D_{A_1}} \omega_{A_1} \otimes \sigma (\omega_{A_1}) P_1(\omega_{A_1}) 
d\omega_{A_1}
\end{equation}
where $\sigma (\omega_{A_1})$ is a \emph{unique} state in
$\bigotimes_{i=2}^{N} {\cal H}_{A_i}$, and $P_1(\omega_{A_1})$ is a
\emph{unique} probability density on the space of states $D_{A_1}$.
Our strategy will be to construct a family of LSEs for the states
$\sigma(\omega_{A_1})$ and invoke the inductive hypothesis to conclude
that they are separable.

Now, consider the state in ${\cal H}_{A_1} \otimes
(\bigotimes_{i=2}^{N-1} {\cal H}_{A_i}^ {\otimes k}) \otimes {\cal
  H}_{A_N} $ defined by
\begin{equation}
\rho_{(1,k,\ldots,k,1)} = \mathrm{Tr}_{A_1^{\otimes (k-1)}} [\rho_{\vec{n}_k}].
\end{equation}
Note that $\rho_{\vec{n}_k}$ exists by hypothesis but need not be
unique. We will impose the further condition on $\rho_{\vec{n}_k}$ that
for all $m>k$ there is an LSE $\rho_{\vec{n}_m}$ for $\rho$ such that
\begin{equation}
\label{biggerext}
\mathrm{Tr}_{\{A_1^{\otimes m-k} \cdots A_{N-1}^{\otimes m-k}\} } 
[ \rho_{\vec{n}_m}]=\rho_{\vec{n}_k}.
\end{equation}
Thus $\rho_{\vec{n}_k}$ itself has symmetric extensions to larger
numbers of copies of the different parties \footnote{It may not be immediately
obvious that this is possible. Suppose not, then for all $\vec{n}_k$
LSE $\rho_{\vec{n}_k}$ there is some $m>k$
such that there is no
$\vec{n}_m$ LSE satisfying (\ref{biggerext}). Take any $\vec{n}_m$ LSE
$\rho_{\vec{n}_m}$ and consider the state $\mathrm{Tr}_{\{A_1^{\otimes
    m-k} \cdots A_{N-1}^{\otimes m-k}\} } [ \rho_{\vec{n_m}}]$; it is
clear that this state is a $\vec{n}_k$ LSE for $\rho$ which is a
contradiction.}.

It is not difficult to see that if we consider again the bipartite
split $A_1 - (A_2,\ldots,A_{N})$, the state $\rho_{(1,k,\ldots,k,1)}$
has symmetric extensions to any number of copies of ${\cal H}_{A_1}$.
For example, if we want a symmetric extension to $m$ copies, $m\leq
k$, we can take
\begin{equation}
\mathrm{Tr}_{A_1^{\otimes (k-m)}} [\rho_{\vec{n}_k}],
\end{equation} 
(where $\mathrm{Tr}_{A_1^{\otimes 0}}$ means not taking any trace),
and if $m>k$ we take
\begin{equation}
\mathrm{Tr}_{\{A_2^{\otimes m-k} \cdots A_{N-1}^{\otimes m-k}\} } 
[ \rho_{\vec{n}_m}].
\end{equation}
Thus, we have that $\rho_{(1,k,\ldots,k,1)}$ has
symmetric extensions to any number of copies of ${\cal H}_{A_1}$, so
applying the bipartite result again we can write
\begin{eqnarray}
\label{rhoext}
\rho_{(1,k,\ldots,k,1)} & = & \int_{D_{A_1}} \omega_{A_1} \otimes 
\sigma_{(k,\ldots,k,1)}
 (\omega_{A_1}) \times \nonumber \\
 & & \times \  P_{(k,\ldots,k,1)} (\omega_{A_1}) \ d\omega_{A_1},
\end{eqnarray}
where both the state $\sigma_{(k,\ldots,k,1)} (\omega_{A_1})$ in
$(\bigotimes_{i=2}^{N-1} {\cal H}_{A_i}^ {\otimes k}) \otimes {\cal
  H}_{A_N}$ and the probability density $P_{(n_2,\ldots,n_{N-1},1)}
(\omega_{A_1})$ defined on $D_{A_1}$ are \emph{unique}.

If we trace out $(k-1)$ copies of ${\cal H}_{A_i}$, $2\leq i \leq
N-1$, in (\ref{rhoext}), we obtain
\begin{eqnarray}
\label{rho2}
\rho & = & \int_{D_{A_1}} \omega_{A_1} \otimes \mathrm{Tr}_{\{A_2^{\otimes k-1} 
\cdots A_{N-1}^{\otimes k-1}\} } 
[\sigma_{(k,\ldots,k,1)}
 (\omega_{A_1})] \times \nonumber \\
& & \times \  P_{(k,\ldots,k,1)} (\omega_{A_1}) \ d\omega_{A_1}.
\end{eqnarray}
If we compare (\ref{rho1}) and (\ref{rho2}), we can use the uniqueness
of the decomposition to conclude that
\begin{equation}
\label{ext}
\sigma (\omega_{A_1}) = \mathrm{Tr}_{\{A_2^{\otimes k-1} 
\cdots A_{N-1}^{\otimes k-1}\} } 
[\sigma_{(k,\ldots,k,1)}
 (\omega_{A_1})],\,\,\, \forall k\geq 1,
\end{equation}
and
\begin{equation}
 P_{(k,\ldots,k,1)} (\omega_{A_1}) =  P_1(\omega_{A_1}). 
\end{equation}
For each $\omega_{A_1}$, the state $\sigma (\omega_{A_1})$ is a state
in $\bigotimes_{i=2}^{N} {\cal H}_{A_i}$. We claim that this state has
locally symmetric extensions for the first $(N-2)$ parties that are
associated with vectors of $N-1$ components of the form $\vec{n}_k =
(k,k,\ldots,k,1), \, \forall k\geq 1$.

Equation (\ref{ext}) proves the existence of the extensions. To prove
the symmetry, we use equation (\ref{rhoext}) and uniqueness of the
decomposition. First note that, by hypothesis, we can state that
\begin{equation}
\label{restrsym}
\rho_{(1,k,\ldots,k,1)} = V_{i,\tau (i)} \, \rho_{(1,k,\ldots,k,1)} V_{i,\tau (i)},
\end{equation}
which holds $\forall i,\ 2\leq i \leq (N-1)$, and $\forall \tau (i)
\in S_k$, since these symmetry requirements are implied by the
symmetry properties of $\rho_{\vec{n}_k}$. Note that the permutation
operators in (\ref{restrsym}) act only on parties $A_2$ through
$A_{N-1}$. If we apply (\ref{restrsym}) to both sides of
(\ref{rhoext}), we obtain
\begin{eqnarray}
\label{rhoext2}
\rho_{(1,k,\ldots,k,1)} & = & \int_{D_{A_1}} \omega_{A_1} \otimes \left( V_{i,\tau (i)}
\ \sigma_{(k,\ldots,k,1)}
 (\omega_{A_1})\  V_{i,\tau (i)}\right)  \nonumber \\
& & \times \   P_1 (\omega_{A_1}) \ d\omega_{A_1}.
\end{eqnarray} 
But comparing (\ref{rhoext2}) with (\ref{rhoext}), and using again the
uniqueness of the decomposition, we have
\begin{equation}
 \sigma_{(k,\ldots,k,1)}(\omega_{A_1}) = V_{i,\tau (i)}
\ \sigma_{(k,\ldots,k,1)}
 (\omega_{A_1})\  V_{i,\tau (i)}.
\end{equation}
So the extensions of $\sigma (\omega_{A_1})$ have the required
symmetry.

We can now apply the inductive hypothesis to $\sigma(\omega_{A_1})$
and conclude that this state must be fully separable and in fact
\begin{eqnarray}
\label{expans}
\sigma (\omega_{A_1}) & = &  \int_{D_2^{N-1}}
\left(\bigotimes_{i=2}^{N-1} \omega_{A_i}\right) \otimes 
\lambda_{A_{N}} (\omega_{A_1},\ldots,\omega_{A_{N-1}}) \nonumber \\
& & \times \ \Pi_{i=2}^{N-1} P_{i}(\omega_{A_{i}} | \omega_{A_{i-1}},\ldots,\omega_{A_1})
\ d\omega_{A_i}. 
\end{eqnarray}
Combining (\ref{rho1}) with (\ref{expans}) we finally get
\begin{eqnarray}
\rho & = &  \int_{D_1^{N-1}} \left( \bigotimes_{i=1}^{N-1}\omega_{A_i}
\right) \otimes 
\lambda_{A_{N}} (\omega_{A_1},\ldots,\omega_{A_{N-1}})  \nonumber \\
& & \times \ \Pi_{i=1}^{N-1} P_{i}(\omega_{A_{i}} | \omega_{A_{i-1}},\ldots,\omega_{A_1})
\ d\omega_{A_i},
\end{eqnarray} 
showing that the state $\rho$ is fully separable. $\Box$.

This result generates a sequence of separability criteria labeled by
the integer $k$. Since the existence of a LSE for some $k_1$ implies
the existence of a LSE for all $k_2$, $k_2 \leq k_1$, then we have
that this sequence has a hierarchichal structure, similar to the one
introduced for the bipartite case in~\cite{doherty2004a}. We note that
this sequence of state extensions is exactly the one considered
in~\cite{wolf2003a} in context of finding local hidden variable
theories for multipartite states $\rho$. This shows that, exactly as
in \cite{terhal2003a}, these local hidden variable theories can only
give a local realistic description of Bell experiments having an
arbitrary number of detector settings for the two observers when the
states of interest are separable.  However, applying this particular
hierarchy of tests is not the best practical tool to detect
entanglement of multipartite states.

>From Theorem 1 we have the following corollary:
\newtheorem{thm2}{Corollary}
\begin{thm2}
  A multipartite mixed state $\rho$ in $\bigotimes_{i=1}^N {\cal
    H}_{A_i}$ has PPTLSE to any number of copies of its first $(N-1)$
  parties, if and only if, $\rho$ is fully separable.
\end{thm2}
\textbf{Proof:} If $\rho$ is fully separable, it has a decomposition
of the form (\ref{sep}) and hence we can construct the PPTLSE given by
(\ref{lse}). On the other hand, if $\rho$ has PPTLSE to any number of
copies of its first $(N-1)$ parties, in particular it has PPTLSE to
extensions associated with the vectors $\vec{n_k} = (k,k,\ldots,k,1),
\, \forall k\geq 1$. Since any PPTLSE is also a LSE, according to
Theorem 1 $\rho$ must be fully separable. $\Box$ (Note that we could
replace PPTLSE by LSE in the statement of Corollary 1 and still
recover the same result).

Corollary 1, although equivalent to Theorem 1, seems to be less
practical, since we require the existence of many more PPTLSE.
However, since the existence of any PPTLSE is a \emph{necessary}
condition for separability, its nonexistence is a \emph{sufficient}
condition for entanglement.  The advantage of an application of these
results based on Corollary 1 rather than on Theorem 1, lies in the
fact that we might be able to show entanglement by searching for a
PPTLSE to one extra copy of \emph{one} of the parties instead of one
extra copy of \emph{all} parties. In terms of the resources needed to
implement this might amount to a huge saving. For example, if we have
a state in $2 \otimes 4 \otimes 4$, it is much easier to search for a
PPTLSE to 3 copies of the first party, than it would be to search for
a PPTLSE to one copy of each of the parties. Corollary 1 gives us the
chance of choosing a more economical way of testing for entanglement.
We will see later on, when we discuss a particular example, that this
approach can be very useful.

In \cite{horodecki2000b} the multipartite separability problem was
discussed in terms of linear maps positive on products states. Every
multipartite entanglement witness can be transformed into such a
linear map and our result has implications for the characterization of
these maps. In \cite{doherty2004a} we characterised strictly positive
maps as those that are completely positive when composed with one of a
class of maps onto the symmetric subspace of some number of copies of
the output space of the linear map. An exactly similar
characterization of the adjoint of a linear map strictly positive on
product states is possible based on Theorem 1. Since the only
difference is extending notation of \cite{doherty2004a} to the
multipartite case we will not give an explicit discussion.

\section{Example}
\label{example}
Here we consider the example of a complete bound entangled three qubit
state constructed by Bennett {\it et al.} from an unextendible product
basis \cite{bennett1999a}.


In the example, we look for one-copy extensions of one of the parties,
i.e., the case where $\vec{n} = (2,1,1)$. Equivalently, from the dual
viewpoint, we look for witnesses $Z$ for which $ |x|^2 \langle x y z |
Z| x y z \rangle$ has a decomposition as a sum of squares magnitudes.


\subsection{A $2 \otimes 2 \otimes 2$ state from UPBs}

We apply the results to a $2 \otimes 2 \otimes 2$ tripartite state,
first proposed in \cite{bennett1999a}. This entangled state is
constructed using unextendible product bases (UPBs), and has the very
interesting property of being separable for every possible bipartition
of the three parties. The state has the following expression:
\begin{equation}
\rho = \frac{1}{4}
(\mathbf{1} - \sum_{j=1}^4 |\psi_j \rangle \langle \psi_j|),
\label{eq:upbstate}
\end{equation}
where 
\begin{align*}
\psi_1 = | 0,1,+ \rangle, \qquad
&\psi_2 = | 1,+,0 \rangle, \\
\psi_3 = | +,0,1 \rangle, \qquad
&\psi_4 = | -,-,- \rangle,  
\end{align*}
and $\pm = (|0 \rangle \pm |1 \rangle) / \sqrt{2}$.  After solving the
SDP, we easily arrive at a witness whose matrix representation is
given below:
\begin{equation}
Z = 
\left[
\begin{array}{rrrrrrrr}
     1 &  -1 &   -1 &    1 &   -1 &    1 &    1 &   -1 \\
    -1 &   4 &    1 &    0 &    1 &    3 &   -1 &    1 \\
    -1 &   1 &    4 &    3 &    1 &   -1 &    0 &    1 \\
     1 &   0 &    3 &    4 &   -1 &    1 &    1 &   -1 \\
    -1 &   1 &    1 &   -1 &    4 &    0 &    3 &    1 \\
     1 &   3 &   -1 &    1 &    0 &    4 &    1 &   -1 \\
     1 &  -1 &    0 &    1 &    3 &    1 &    4 &   -1 \\
    -1 &   1 &    1 &   -1 &    1 &   -1 &   -1 &    1 
\end{array}
\right].
\end{equation}

It can be verified that $\mathrm{Tr} [Z \rho] = -\frac{3}{8} < 0$, but
$Z$ is nonnegative in all product states. This is certified by an
identity, obtained from the solution of the SDP, that
expresses $ |x|^2 \langle x y z | Z | x y z \rangle$ as a sum of
squared magnitudes.

\section{Conclusions}
\label{conclude}

In this paper we have discussed separability criteria for multipartite quantum
states based on the
existence of extensions of the state to a larger space consisting of
several copies of each of the subsystems. The symmetric extensions we
consider always exist if the state is separable but do not necessarily
exist for entangled states. We showed that multipartite entangled
states will eventually fail one of these tests and in this case we constructed
an entanglement witness using the duality theory of semidefinite
programming. 

It would be enlightening to better understand the physical
significance of these symmetric extensions. It can be said that they
highlight a version 
of the monogamy of 
entanglement for mixed states; if a group of entangled systems are in
a strongly
entangled state it is hard for them to share the same entanglement
with other systems. Another interpretation carries over from the
bipartite case, the
symmetric state extensions, if they exist, provide local hidden
variable descriptions for large classes of possible multiparty Bell
experiments. This is discussed much more fully in
\cite{wolf2003a,terhal2003a}. 

Other questions for further study include the behavior of our tests under local
operations and classical communication. Unlike the positive partial
transpose test it is not clear that the property of having a symmetric
extension to a given number of copies of the subsystems is preserved
under local operations and classical communication. Certainly the
tests we construct are invariant under local unitary operations but,
just as in the bipartite case~\cite{doherty2004a}, there are state
transformations that may be achieved with some probability by local
operations and classical communication that can convert a state having
a symmetric state extension into one that does
not. A sequence of tests for entanglement that was invariant under
local operations and classical communication would point to the
existence of many sets of states, other than the positive partial
transpose states and the separable states, that are closed under local
operations and classical communication and this could have interesting
consequences for quantum information theory. Another important
open question is the problem of finding explicit product state
decompositions for separable states. As they exist at the moment our
tests only provide definitive answers when the state of interest turns
out to be entangled. A more powerful procedure would be able to
detect separable states and construct product state decompositions
when this is possible.

\section*{Acknowledgements}
ACD and FMS acknowledge support from the NSF Institute for Quantum
Information under grant No. EIA-0086083, and FMS is grateful for
support from the National Research Council. ACD was also supported in
part by the Caltech MURI Center for Quantum Networks
(DAAD19-00-1-0374).
We would like to thank Jens Eisert for letting us know about
closely related work that was simultaneously posted on the quant-ph archive
\cite{eisert2004a}.

\bibliographystyle{apsrev} \bibliography{Cav1}

\end{document}